\documentclass{aa}  

\usepackage{graphicx}
\usepackage{txfonts}
\begin{document}

\title{DN Tauri - coronal activity and accretion in a young low-mass CTTS}
   \author{J. Robrade\inst{1}
          \and M. G{\"u}del\inst{2}
          \and H.M. G{\"u}nther\inst{3}
         \and J.H.M.M. Schmitt\inst{1}
          }
   \institute{Hamburger Sternwarte, University of Hamburg, Gojenbergsweg 112, 21029 Hamburg, Germany
   \and
   Department of Astronomy, University of Vienna, T{\"u}rkenschanzstr. 17, 1180 Vienna, Austria
   \and
   Harvard-Smithsonian Center for Astrophysics, 60 Garden Street, Cambridge, MA 02138, USA\\
       \email{jrobrade@hs.uni-hamburg.de} 
  }

\date{received: 24/09/2013, accepted: 14/11/2013}
  \abstract
 {Low-mass stars are known to exhibit strong X-ray emission during their early evolutionary stages.
This also applies to classical T Tauri stars (CTTS), whose X-ray emission differs from that of main-sequence stars in a number of aspects.}
 {We study the specific case of DN Tau, a young M0-type accreting CTTS, to extend the range of young CTTS studied with high-resolution X-ray spectroscopy at lower masses
and to compare its high-energy properties with those of similar objects.}
  {We use a deep XMM-Newton observation of DN Tau to investigate its X-ray properties and X-ray generating mechanisms;
specifically we examine the presence of
X-ray emission from magnetic activity and accretion shocks. We also compare our new X-ray data with UV data taken simultaneously and with X-ray/UV observations performed before.
}
{We find that the X-ray emission from DN~Tau is dominated by coronal plasma generated via magnetic activity, but also clearly detect a contribution of the accretion shocks to the cool plasma component at $\lesssim 2$~MK
as consistently inferred from density and temperature analysis. Typical phenomena of active coronae like flaring, the presence of very hot plasma at 30~MK and
an abundance pattern showing the inverse FIP effect are seen on DN~Tau.
Strong variations in the emission measure of the cooler plasma components between the 2005 and 2010 data point
to accretion related changes; in contrast the hotter coronal plasma component is virtually unchanged. The UV light curve
taken simultaneously is in general not related to the X-ray brightness, but exhibits clear counterparts during the observed X-ray flares.  
}
   {The X-ray properties of DN~Tau are similar to those of more massive CTTS, but its low mass and large radius associated with its youth shift the 
   accretion shocks to lower temperatures, reducing their imprint in the X-ray regime. DN~Tau's overall X-ray properties are dominated by strong magnetic activity.
}


   \maketitle
%

\section{Introduction}

Classical T~Tauri stars (CTTS) are pre-main sequence, low mass stars (M~$\lesssim $~2\,M$_{\odot}$) that are 
still accreting matter from a surrounding circumstellar disk at a significant level. 
CTTS show an emission line spectrum and are often characterized by strong veiling and a large H$\alpha$ equivalent width, formerly the main classifier of this class.
The H$\alpha$ equivalent width criterion for a CTTS is, however, spectral-type
dependent. Further studies have shown that the 10\,\% width of the H$\alpha$ line
is mostly determined by the accretion streams and is less spectral-type dependent and a more reliable tracer of accretion, especially in low mass stars \citep[see e.g.][]{whi03}.
Beside H$\alpha$ many other lines have been used to quantitatively study accretion properties in young stars.
In addition to emission lines, CTTS show a strong IR-excess from their circumstellar disk that also provides a diagnostic on the inclination of the system.  
CTTS evolve via transitional objects, where the disk starts to dissipate and accretion rates decrease, into
weak-line T~Tauri stars (WTTS) that are virtually disk-less and do not show strong signs of ongoing accretion.

T~Tauri stars are known to be strong and variable X-ray emitters from
{\it Einstein} and {\it ROSAT} observations.
X-ray studies of star forming regions such as COUP \citep[{\it Chandra} Orion Ultradeep Project,][]{get05}
or XEST \citep[{\it XMM-Newton} Extended Survey of the Taurus Molecular Cloud,][]{gue07},
confirmed that T~Tauri stars show high levels of magnetic activity as evidenced by hot coronal plasma and strong flaring,
but also refined the general X-ray picture of YSOs (Young Stellar Objects).

In the commonly accepted magnetospheric accretion model for CTTS \citep[e.g.][]{koe91} material is accreted from the stellar disk onto the star
along magnetic field lines, which disrupt the accretion disk in the vicinity of the corotation radius. 
Since the infalling material originates from the disc truncation radius, typically located at several stellar radii, 
it reaches almost free-fall velocity and upon impact the supersonic flow forms a strong shock near the stellar surface.
The funnelling of the accreted matter by the magnetic field leads to the formation of accretion spots that have small surface filling factors \citep{cal98}
and produce strong optical/UV and X-ray emission \citep{lam98, guenther07}.

The X-ray emission from accretion spots on CTTS has specific signatures which are detectable with high-resolution X-ray spectroscopy.
Accretion shocks generate plasma with temperatures of up to a few MK that is significantly cooler than the average coronal plasma
and sufficiently funneled accretion streams are expected to produce X-rays in a high-density environment ($n_{e} \gtrsim 10^{11}$\,cm$^{-3}$ measured in \ion{O}{vii}).
In contrast, plasma produced by magnetic activity covers a much broader temperature range spanning in total 1\,--\,100~MK and is on average hotter \citep[$T_{\rm av} \approx 10 - 20$~MK
for CTTS in Taurus,][]{tel07}. Further 
it typically has, at least outside large flares, much lower densities \citep[$n_{e} \lesssim 3\times 10^{10}$\,cm$^{-3}$,][]{ness04}.
The accretion streams may also influence coronal structures on the stellar surface or lead to additional magnetic activity via star-disk interaction.
Moreover, the accretion process is
accompanied by outflows or winds from the star and the surrounding disk, which play an important role in star formation via the transport of angular momentum.
Stellar jets and associated shocks provide another X-ray production mechanism, which typically generates cool plasma at low densities,
as seen in several T~Tauri stars such as DG~Tau \citep{gue05}. X-ray diagnostics like
density and temperature sensitive X-ray line ratios can be utilized to distinguish between the different scenarios.

TW~Hya was the first and is still the most prominent CTTS that is dominated by accretion shocks in X-rays. Its X-ray spectrum shows
high density plasma as evidenced by density sensitive lines in He-like triplets of oxygen and neon
and an unusually cool plasma distribution \citep{kas02,ste04}. TW~Hya has been extensively studied in X-rays and a very deep {\it Chandra} observation suggests that
the accreted and shock-heated material mixes with surrounding coronal material, likely
producing a complex distribution of emission regions around the accretion spots \citep{bri10}. 
So far all low-mass CTTS studied at X-ray energies show similar signs of accretion plasma, classic examples are e.g. BP~Tau \citep{schmitt05},
V4046~Sgr \citep{guenther06,arg12}, MP~Mus \citep{arg07} or RU~Lup \citep{rob07b}.
In contrast, T~Tau itself shows a strong cool plasma component, but a low plasma density \citep{gue07a};
although the system is dominated by the intermediate mass T~Tauri star T~Tau~N ($M \approx 2.4~M_{\odot}$).
In a comparative study of several bright CTTS it was shown that the presence of X-rays from both accretion shocks and magnetic activity is likely universal, but that the
respective contributions differ significantly between the individual objects \citep{rob06}.
Indeed, magnetic activity produces the bulk of the observed X-ray emission in the majority of CTTS in the 0.2\,--\,2.0~keV band and completely dominates 
at higher energies.
In addition, X-ray temperature diagnostics have shown that all accreting stars exhibit an excess of shock-generated cooler plasma, 
leading to a soft excess when compared to coronal sources \citep{rob07b, tel07, gue07b}. 
The observed X-ray spectrum of young stars is modified by often significant absorption by circumstellar or disk material as well as by outflowing and infalling matter.
X-ray absorption can exceed optical extinction by an order of magnitude as shown e.g. for RU~Lup \citep{rob07b} and 
may even be strongly time variable as in AA~Tau \citep{schmitt07}.
The different stellar properties such as mass, rotation and activity, varying mass accretion rates and degree of funneling as well as the
viewing angle dependence naturally lead to the variety of X-ray phenomena in YSOs that are an inter-mixture of magnetic activity, accretion and outflow processes.

High resolution X-ray spectra from young accreting stars are only in a few cases available and existing studies
focussed on the more massive CTTS with spectral type G or K.
Young low-mass stars with M~$\lesssim$~0.5\,M$_{\odot}$ are typically X-ray fainter and so far only poorly studied; 
the transitional multiple system \hbox{Hen 3-600} \citep{huene07} is one of the rare examples. 
Nevertheless, they are the most common stars and their investigation is of
great astrophysical interest to draw a more complete and general picture of the evolution of young stars and their surrounding environment, where
the stellar winds and UV/X-ray emission influence the chemistry and evolution of the circumstellar disk and the process of planet formation.

\section{The target: DN Tau}
\label{tar}

Our target star \object{DN Tau} is a M0-type CTTS located in the Taurus Molecular Cloud (TMC) at a distance of $d=140$~pc \citep{coh79};
important stellar parameters collected from the literature are summarized in Table\,\ref{pro}.
DN~Tau is a single star on a fully convective track with an estimated age in the range of 0.5\,--\,1.7~Myr.
Its optical extinction is quite low, indicating that DN~Tau is not deeply embedded in circumstellar material or the TMC.
While classical estimates of stellar luminosity, mass and radius are about $L_{*}= 1.0~L_{\odot}$, $M_{*}= 0.5~M_{\odot}$ and $R_{*}= 2.1~R_{\odot}$, 
\cite{donati13} find a slightly hotter, smaller and less luminous
model of DN~Tau ($L_{*}= 0.8~L_{\odot}$, $M_{*}= 0.65~M_{\odot}$, $R_{*}= 1.9~R_{\odot}$) adopting an optically measured $A_{V} =0.5$. In contrast,
\cite{ing13} find a much brighter and larger DN~Tau ($L_{*}= 1.5~L_{\odot}$, $M_{*}= 0.6~M_{\odot}$, $R_{*}= 2.8~R_{\odot}$) when using $A_{V} =0.9$, that bases on $A_{J}=0.29$ from IR-measurements \citep{fur11}.
The CTTS nature of DN~Tau is reflected by a typical EW\,[H$\alpha$]~$ = 12 - 18$~{\r A} and a H$\alpha$ 10\,\% width 
in the range of $290 - 340$~km\,s$^{-1}$ \citep{herb88, whi04,ngu09} and a moderate IR excess, 
making it a Class II source based on its far-IR SED \citep{ken95}.
DN~Tau is a variable, but typically moderate accretor that exhibits little UV excess;
e.g. \cite{gull98} derived $L_{\rm acc} = 0.016~L_{\odot}$ and a weak optical veiling of $r=0.075$. 
Nevertheless, the infalling plasma on DN~Tau is apparently well funnelled with an accretion spot filling factor of $f=0.005$ \citep{cal98}.
\cite{ing13} modeled broad-band optical and UV data in a similar approach but with multiple accretion columns and found,
depending on the absence/presence of 'hidden' low flux accretion emission, $f=0.002/0.06$ and $\log \dot{M}_{\rm acc} = -8/-7.8~M_{\odot}$\,y$r^{-1}$.
\cite{donati13} give $\log \dot{M}_{\rm acc} = -9.1 \pm 0.3~M_{\odot}$\,yr$^{-1}$ as average for their accretion proxies.
While the reliability of the various methods used to obtain quantitative estimates on the mass accretion rates is debated,
highly variable accretion properties of DN~Tau are observed and
\cite{fern95} measured an EW\,[H$\alpha$] declining from 87 to 15~{\r A} within four days.
DN~Tau has only a weak outflow; while typically about 5\,--\,10\,\% of the accretion rate are estimated, \cite{whi04} give a 2\,\% upper limit derived from their data.

\begin{table}[t]
\begin{center}
\caption{\label{pro}Stellar properties of DN Tau from optical measurements.}
\begin{tabular}{lcr}
\hline\hline\\[-3mm]
 Sp. type & M\,0\,$^{1,2,3}$  & \\
$T_{\rm eff}$ & 3800\,$^{3}$ ... 3850\,$^{1}$ ... 3950$\pm 50$\,$^{4}$ & K\\
$M_{*}$&   0.4\,$^{1,2}$ ... 0.5\,$^{3}$ ... 0.65$\pm 0.05$\,$^{4}$& $M_{\odot}$\\
$R_{*}$&  1.9$\pm 0.2$\,$^{4}$ ... 2.1\,$^{2}$ & $R_{\odot}$\\
$L_{\rm bol}$ & 0.8$\pm 0.2$\,$^{4}$ ... 0.9\,$^{2}$ ... 1.0\,$^{1}$ &  $L_{\odot}$\\
$A_{V}$ & 0.25\,$^{2}$ ... 0.5\,$^{1}$& mag\\
$\log \dot{M}_{\rm acc}$ & -7.8\,$^{3}$ ... -8.5\,$^{2}$ ... -9.1$\pm 0.3$\,$^{4}$  & $M_{\odot} yr^{-1}$\\\hline
\end{tabular}
\end{center}
\noindent
{\scriptsize $^{1}$ \cite{ken95}, $^{2}$ \cite{gull98}, $^{3}$ \cite{whi04}, \\$^{4}$ \cite{donati13}}
\end{table}

Photometric variations of DN~Tau's optical brightness were first reported with a period of about $P_{\rm rot} \approx 6.0$~d \citep{bou86},
later refined to $P_{\rm rot} = 6.3$~d \citep{vrba93}. This variability can be interpreted as rotational modulation 
of a large magnetic spot or spot group with a surface coverage of up to 35\,\%. 
Strong magnetic activity on DN~Tau is also implied by its large inferred mean magnetic field of 2~kG \citep{joh07}.
Results from spectropolarimetric observations with ESPaDOnS/CFHT \citep{donati13}
show a simple magnetic topology that is largely axisymmetric and mostly poloidal with a dominant octupolar and a weaker bipolar component of 0.6\,--\,0.8.~kG and 0.3\,--\,0.5~kG polar strength respectively.
\cite{muz03} present near-IR spectra of DN~Tau from which they inferred
an inner (dust-)disk rim located at 0.07~AU ($\approx 7~R_{*}$), notably the closest disk rim in their sample. 
The disk of DN~Tau with $M_{d} = 0.03~M_{\odot}$ as deduced from submillimeter observations is quite massive, roughly an order of magnitude above the median mass found for the Class~II sources in the sample of \cite{and05}.
DN~Tau is viewed under an intermediate inclination;
\cite{muz03} inferred an inclination of $i=28\pm 10^{\circ}$ from IR data, quite similar to the estimate of $i=35\pm 10^{\circ}$ by \cite{donati13}.
Adopting $P_{\rm rot} = 6.3$~d,  $i=33^{\circ}$ and combining these data with the rotational velocity of
$v$\,sin\,$i= 12.3 \pm 0.6$~km\,s$^{-1}$ \citep{ngu09}, $v$\,sin\,$i = 9 \pm 1$~km\,s$^{-1}$ \citep{donati13} or $v$\,sin\,$i= 10.2$~km\,s$^{-1}$ \cite{hart89},
we obtain $R_{*} \approx 2.8~R_{\odot}$,  $R_{*} \approx 2.0~R_{\odot}$ and $R_{*} \approx 2.3~R_{\odot}$ respectively.

X-ray emission from DN~Tau was first detected with {\it Einstein} \citep{wal81} and later by {\it ROSAT} \citep{neu95}; both at a similar X-ray luminosity of $\log L_{\rm X} \approx 29.7$~erg\,s$^{-1}$, albeit with significant error.
DN~Tau has been observed by {\it XMM-Newton} in 2005 as part of the XEST project (No. 12-040); an analysis of these data is presented in \cite{tel07}.
They derived basic X-ray properties from an EMD model and a multi-temperature fit, both methods give $L_{\rm X} = 1.2 \times 10^{30}$~erg\,s$^{-1}$ and an average coronal temperature of about 12\,--\,14~MK.
DN~Tau is among the X-ray brighter CTTS, when compared to similar objects in the XEST or COUP sample and its X-ray activity level is with $\log L_{\rm X}/L_{\rm bol} \approx -3.5$ close to, 
but still about a factor of three below, the saturation limit at $\log L_{\rm X}/L_{\rm bol} \approx -3$.
We re-observed DN~Tau in 2010 with {\it XMM-Newton}, primarily to obtain a deeper exposed
high-resolution X-ray spectrum with the aim to expand the sample of emission line studied CTTS into the lower mass regime.

In this paper we present an analysis of the new {\it XMM-Newton} observations of DN~Tau
and compare it to earlier observations.
Our paper is structured as follows: in Sect.\,\ref{obsana} the X-ray observations and the data analysis are described,
in Sect.\,\ref{results} we present our results subdivided into different physical topics, in Sect.\,\ref{comp} we 
discuss our DN~Tau results and compare it to other CTTS and end with a summary in Sect.\,\ref{sum}.

\begin{table}[t]
\begin{center}
\caption{\label{log} XMM-Newton observing log of DN Tau.}
\begin{tabular}{rrrr}\hline
Date & Obs. ID.  & Dur. MOS/PN (ks)  \\\hline\\[-3.mm]
2005-03-04/05& 0203542101 & 31/29 \\
2010-08-18/19& 0651120101 & 119/118 \\\hline
\end{tabular}
\end{center}
\end{table}

\section{Observations and data analysis}
\label{obsana}

The target DN Tau was observed by {\it XMM-Newton} twice; a 30~ks exposure was obtained for the XEST survey in 2005 (PI: Guedel)
and a 120~ks exposure was obtained in 2010 (PI: Robrade). We focus on the deeper 2010 exposure, 
but also re-analyze the 2005 data in an identical fashion to ensure consistency throughout this work.
Data were taken with all X-ray detectors, i.e. the EPIC (European Photon Imaging Camera)
and the RGS (Reflection Grating Spectrometer) as well as the optical monitor (OM). The EPIC consists of two MOS and one PN detector; the PN is the more sensitive instrument
whereas the MOS detectors have a slightly higher spectral resolution.
The EPIC instruments were operated in both observations in the full frame mode with the medium filter, allowing a direct comparison of the data.
The OM was operated in the fast mode with the U filter in 2005 (eff. wavelength 3440 \AA) and the UVW1 (eff. wavelength 2910 \AA) filter in 2010.
A detailed description of the instruments can be found in the 'XMM-Newton Users Handbook' (http://xmm.esac.esa.int); the used data is summarized in Table\,\ref{log}.

All data analysis was carried out with the {\it XMM-Newton} Science Analysis System (SAS) version~11.0 \citep{sas} and standard SAS tools were used
to produce images, light curves and spectra.
Standard selection criteria were applied to the data, light curves are background subtracted and we exclude periods of high background from spectral analysis.
Source photons from the EPIC detectors were extracted from circular regions around DN~Tau and the background was taken from nearby source free regions. 
The RGS data of DN~Tau has only a moderate SNR, therefore we
extracted spectra from a 90\,\%~PSF source region to reduce the background contribution. 
The data of the X-ray detectors are analyzed independently for each observation to study variability and cross-check the results from the different instruments.
We note that some degradation has occurred for the RGS detector between the exposures, while the effective area of the EPIC detectors shows only minor changes.

Spectral analysis was carried out with XSPEC V12.6 \citep{xspec} and we used multi-temperature APEC/VAPEC models \citep{apec} 
with abundances relative to solar photospheric values as given by \cite{grsa} to derive X-ray properties like luminosities or emission measure distributions (EMD). 
We find that photoelectrically absorbed three-temperature models adequately describe the data, but note that some of the fit parameters are mutually dependent, e.g. absolute abundances 
and emission measure, emission measures and temperatures of neighboring components or
absorption column density, temperature and emission measure of cool spectral components. Spectra are re-binned for modeling and
errors in spectral models are given by their 90\% confidence range and were calculated
by allowing variations of normalizations and respective model parameters.
Additional uncertainties may arise from errors in the atomic data and instrumental calibration.
For line fitting purposes we use the CORA program \citep{cora}, identical line widths and Lorentzian line shapes.
Emitted line fluxes are corrected for absorption by using the {\it ismtau}-tool of the PINTofALE software \citep{poa} and flux-conversion is made with the SAS tool {\it rgsfluxer}.

\section{Results}
\label{results}
Here we report on the results obtained from the {\it XMM-Newton} observations, subdivided into separate topics.

\begin{figure}[t]
\includegraphics[width=89mm]{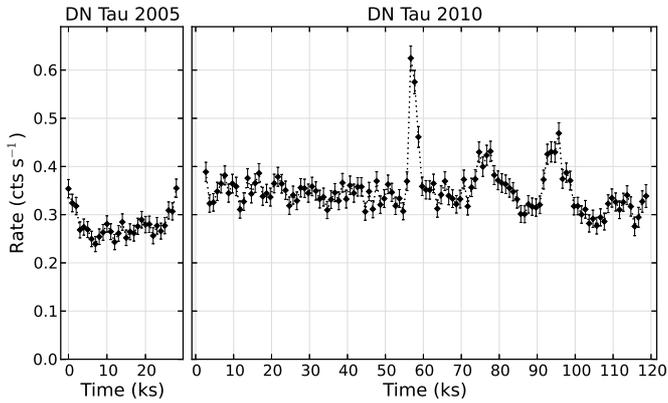}
\caption{\label{lcs} X-ray light curves of DN Tau in 2005 and 2010, 0.2\,--\,5.0~keV EPIC data with 1~ks binning.}
\end{figure}
\begin{figure}[t]
\includegraphics[width=90mm]{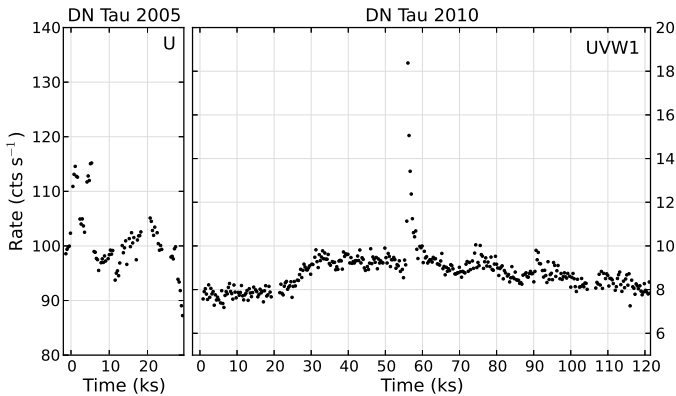}
\caption{\label{omall} OM light curves from 2005 (U filter) and 2010 (UVW1 filter), 300~s binning each.}
\end{figure}

\subsection{X-ray light curves and hardness}

The X-ray light curves of DN~Tau as obtained from the summed EPIC data are shown in Fig.\,\ref{lcs}, here we
use the 0.2\,--\,5.0~keV energy band and a 1~ks temporal binning.
Some variability and minor activity is present in both observations, but large fractions of the X-ray light curves are quite flat and 
only one moderate flare visible at 55\,--\,60~ks with a factor two increase in count rate and two smaller events peaking at about 75~ks and 95~ks are detected during the 2010 exposure.
The features at the beginning and end of the 2005 observation are likely also decay and rise phases of partly covered flares.
Except for a higher average count rate by 20\,--\,30\,\% in 2010, the level of variability within each observation period is comparable.
A long-term trend of declining X-ray count-rate by roughly 10\,\% is seen in the 2010 data and might be
due to rotational modulation. The 1.4~d observation has a phase coverage of about 0.22 and given the moderate inclination of DN~Tau rotational modulation can be expected
for surface features at low and intermediate latitudes.

\begin{figure}[t]
\includegraphics[width=90mm]{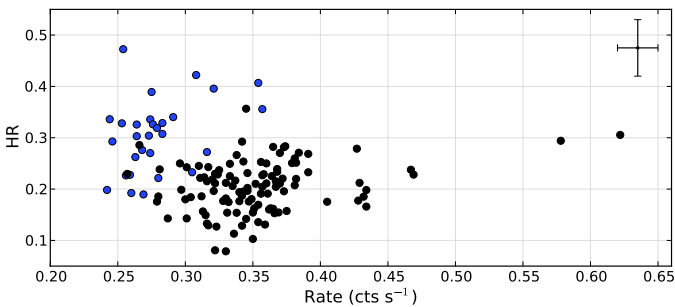}
\caption{\label{hrepic} Hardness ratio of DN Tau from EPIC data, 2010 (black), 2005 (blue); typical errors are indicated in the upper right corner.}
\end{figure}

We investigate the basic spectral state of DN~Tau for both exposures and its evolution with a hardness ratio analysis, $HR=(H-S)/(H+S)$ with
0.2\,--\,0.8~keV as soft band and 0.8\,--\,5.0~keV as hard band.
The energy bands are chosen in a way that X-ray emission in our soft band is predominantly produced by plasma at temperatures of 2\,--\,5~MK, 
whereas the hard band is dominated by emission from hotter plasma at 5\,--\,20~MK; however a moderate shift of the band-separation energy does not influence the results.
As shown in Fig.\,\ref{hrepic} the positive correlation between X-ray brightness and spectral hardness that is typically observed for magnetic activity is generally not present in DN~Tau.
We detect a spectral hardening during the larger flares in 2010, but overall a clear correlation between brightness and hardness is not present. 
Typically more active coronal stars exhibit harder spectra \citep[see e.g.][]{schmitt97}, but a similar trend is also seen when studying the temporal behavior
in individual objects including CTTS \citep{rob06}.
Remarkably, we find that the X-ray fainter state in the year 2005 is overall characterized by harder emission than the brighter 2010 state.
In addition, the individual observation periods again show a broad scatter and only marginal correlations.
Similar conclusions are obtained when inspecting the time evolution of the hardness ratio.

\subsection{UV light curves, flares and UV/X-ray correlations}

The OM light curves of DN Tau are plotted in Fig.\,\ref{omall}; associated brightness
errors are in the range of 0.01\,mag and below the size of the shown symbols.
Note that the DN~Tau observation from 2005 was performed in the U band filter (NUV), while in 2010 the bluer, but less sensitive UVW1 filter (NUV-MUV) was used. 

Comparing the UV light curves with the X-ray ones (Fig.\,\ref{lcs}), it is evident that
the short term variations of the 2005 U band data do not strictly correlate with those of the X-ray brightness, 
suggesting a different origin of the respective emission.
Similarly, \cite{vrba93} find the U band brightness variations to be rather stochastic and not 
modulated by the 6.3~d rotation period as the other optical bands (BVRI). This indicates
that a significant fraction of the UV flux is not associated with the dark spots that are interpreted as magnetically active regions. 
Further, except for a shorter observing period where an out-of-phase modulation was found, a stable accretion configuration, i.e. a dominant hot spot, was not present and overall their U band variations are with $\pm$0.6~mag much larger than those in BVRI with 0.1\,--\,0.2~mag. 
Looking at long-term variations, DN~Tau apparently brightened in the UV-range over the last decades without comparable changes in optical bands.
While there is also significant variability on shorter timescales, the U band brightness increased on average from about $14.5~(14.0-14.9)$~mag during the 
monitoring in the 1980s over $13.9 (14.3-13.3)$~mag 
in the early 1990s \citep{gran07} to a magnitude of $13.25~(13.12-13.36)$~mag (1.5~ks {\it XMM-Newton} average) in 2005.

Also the moderately 'harder' UVW1 flux during the quasi-quiescent part of the 2010 observation is apparently not correlated with X-ray brightness.
During our observation the UV flux varies significantly on timescales of minutes to hours;
for example we observe at about 25~ks an increase of the UVW1 rate by roughly 20\% within a few ks, but without any corresponding X-ray signature
as might be expected for magnetic activity.
Since the photospheric UV emission is negligible in M type stars, the observed behavior favors
a scenario where the bulk of the UV emission is related to several accretion spots located
on the surface of DN~Tau and variability is created by changes in geometry and/or variable spot brightness.

A rough estimate of the relative contributions from magnetic activity and accretion to the UV flux of DN~Tau can be obtained by a comparison
to purely magnetically active sources under the assumption of similar X-ray generating coronae and magnetically induced chromospheric UV emission.
Here we use the active mid-M dwarf EV~Lac \citep{mit05} that was observed with the same instrumental setup as DN~Tau in 2010. 
Accounting for radius and distance, i.e. enlarging EV~Lac to the size of DN~Tau and putting it at the same distance, 
we find that a scaled up version of an active M dwarf outshines DN~Tau by a factor of 1.5\,--\,2.0 in X-rays, but would only produce 15\,--\,20\,\% of its UV flux.
When scaling these values to DN~Tau's true X-ray emission, i.e. accounting for the X-ray overluminosity of the scaled up M dwarf, 
only about 10\,\% of the UV flux from DN~Tau are attributable to magnetic activity.
While a mild suppression in X-ray brightness in accreting vs. non-accreting T Tauri stars is quite typical and might be related to phenomena not present on M dwarfs, 
this comparison shows that the bulk of the UV emission from DN~Tau is generated in the accretion shocks.

\begin{figure}[t]
\includegraphics[width=90mm]{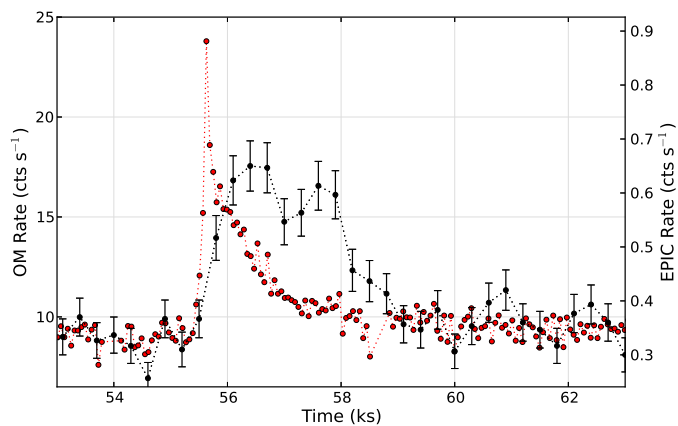}
\caption{\label{om2010}The largest 2010 flare as observed in X-rays (black, 300 s binning) and in the UV (red, 60 s binning.)}
\end{figure}

In contrast, the three X-ray flares observed in 2010 have clear counterparts in the UVW1 data, most prominent is the largest flare starting around 55~ks.
In Fig.\,\ref{om2010} we show its UV light curve overplotted with the X-ray light curve, scaled to the same
quasi-quiescent pre-flare level for clarity. The UV emission precedes the X-rays and peaks about 10~min earlier. 
This behavior indicates an energy release via magnetic reconnection, succeeded by evaporation of fresh material from the stellar surface
that is subsequently heated to X-ray emitting temperatures; flare events like these are frequently observed on the Sun and low-mass stars.
Using results from spectral modelling (see Sect.\,\ref{spec}), we estimate
for the largest flare a peak luminosity of $L_{\rm X} = 3 \times 10^{30}$\,erg\,s$^{-1}$, an energy release of about $2.5 \times 10^{33}$\,erg at X-ray energies and 
a loop length of about $L \approx 0.15~R_{*}$, i.e. an event in a compact coronal structure located close to the stellar surface of DN~Tau.
The time evolution of the flare is dominated by the initial event, but shows sub-structure as visible in the optical plateau 
and the secondary X-ray peak, indicating subsequent magnetic activity. About one hour after the flare onset the X-ray and UV light curves roughly reach their pre-flare values again.

The two smaller X-ray flares show even more complex light curves. They probably result from an overlay of multiple events, for example several magnetic reconnections
occurring within a short time interval in an active region or region complex. Both flares again show UV counterparts, but these are less pronounced than during the large event.

\subsection{Global X-ray properties from CCD spectroscopy}
\label{spec}

\begin{figure}[t]
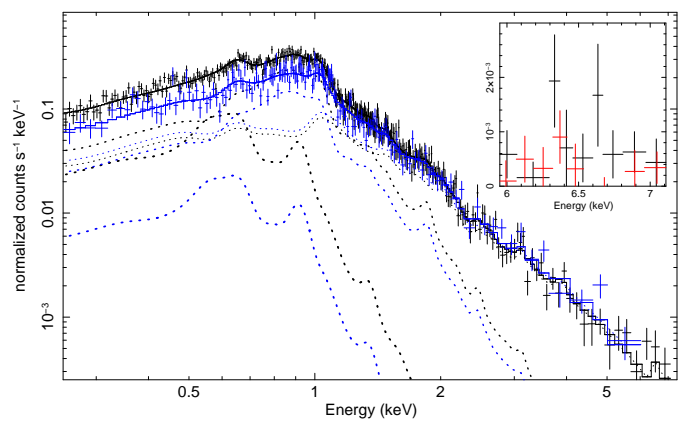

\includegraphics[height=88mm,angle=-90]{dntau_fig5a.ps}

\vspace*{-53.mm}\hspace*{61.5mm}
\includegraphics[height=24mm,angle=-90]{dntau_fig5b.ps}
\vspace*{27mm} 
\caption{\label{pnspec}X-ray spectra of DN Tau ({\it crosses}, PN), spectral fit ({\it histogram}) and model components ({\it dashed})
for the two observations: 2010 (black), 2005 (blue). {\it Inset}: The spectrum above 6.0\,keV during the active (black) and quasi-quiescent (red) half in 2010.}
\end{figure}

To study the global spectral properties of the X-ray emission from DN Tau we use the EPIC data.
As an example of the spectral quality we show in Fig.~\ref{pnspec} the PN spectra and corresponding models for the two observations;
in the inset we show the X-ray emission at high energies which is discussed below.
Visual inspection already shows that major changes have occurred in the softer X-ray regime, whereas the spectra above 1~keV are virtually identical.
The similar spectral slopes suggest that the changes at low energies are not caused by variable line-of-sight absorption, but are intrinsic to the emission of DN~Tau.

The spectral properties of DN~Tau and their changes between the two observations are quantified by modeling 
the spectra in iterative steps to obtain most robust results.
We first investigate potential effects of the flares on the total 2010 spectrum,
but found no differences between the models for the quasi-quiescent ($t < 55$~ks) and active half, except for small re-normalizations by a few percent. 
In a next step we fitted the total data of each observation individually. No significant changes in 
absorption column density and coronal abundances were found between the 2005 and 2010 datasets and we tied these parameters.
We then modeled both observations simultaneously with temperatures and emission measures (EM) as free parameters. 
We first use the MOS detectors that have a better spectral resolution to determine EMDs, abundances and
absorption and cross-checked our results with the PN data, where we also tied the temperatures.
The derived model parameters are given in Table\,\ref{specres} and a
comparison shows that the overall coronal temperature structure and the EMD changes are independent of the data used.
The X-ray luminosities are the emitted ones, i.e. they are absorption corrected; the observed values are given in brackets.
We find for the 2010 (2005) observation X-ray luminosities of $\log L_{\rm X} = 30.2\,(30.1)$~\,erg\,s$^{-1}$,
average coronal temperatures of $T_{\rm X} = 13\,(18)$~MK and an activity level $\log L_{\rm X}/L_{\rm bol} \approx -3.3$,
emphasizing that DN~Tau is among the most active and X-ray brightest CTTS with respect to its mass or effective temperature.

\begin{table}[t]
\caption{\label{specres}Spectral fit results for DN Tau, EPIC data. Parameter absent in the PN results are adopted from the MOS modeling.}
\begin{center}
\begin{tabular}{lrrr}\hline\hline\\[-3.1mm]
Par. & \multicolumn{1}{c}{2005} & \multicolumn{1}{c}{2010} & unit\\\hline\\[-3mm]
& \multicolumn{2}{c}{MOS}&\\\hline\\[-3mm]
$N_{\rm H}$ & \multicolumn{2}{c}{0.8$^{+ 0.1}_{- 0.1}$} &  $10^{21}$cm$^{-2}$\\[1mm]
kT1 &0.17$^{+ 0.05}_{- 0.03}$ &0.23$^{+ 0.03}_{- 0.03}$& keV  \\[1mm]
kT2 &0.60$^{+ 0.07}_{- 0.06}$ &0.64$^{+ 0.03}_{- 0.03}$&  keV\\[1mm]
kT3 & 2.27$^{+ 0.33}_{- 0.21}$ &1.91$^{+ 0.15}_{- 0.14}$&  keV \\[1mm]
EM1 & 0.8$^{+ 0.9}_{- 0.5}$& 2.0$^{+ 0.6}_{- 0.5}$&   $10^{52}$cm$^{-3}$\\[1mm]
EM2 & 3.7$^{+ 0.5}_{- 0.4}$& 5.6$^{+ 0.5}_{- 0.6}$&   $10^{52}$cm$^{-3}$\\[1mm]
EM3 & 6.4$^{+ 0.6}_{- 0.5}$  & 5.3$^{+ 0.5}_{- 0.4}$ & $10^{52}$cm$^{-3}$\\[1mm]
Mg (7.6 eV) & \multicolumn{2}{c}{0.52$^{+ 0.26}_{- 0.18}$}  &  solar \\[1mm]
Fe (7.9 eV) & \multicolumn{2}{c}{0.35$^{+ 0.12}_{- 0.10}$ }&  solar\\[1mm]
Si (8.2 eV)& \multicolumn{2}{c}{0.32$^{+ 0.14}_{- 0.12}$} &  solar\\[1mm]
S (10.4 eV) & \multicolumn{2}{c}{0.24$^{+ 0.18}_{- 0.17}$} &  solar\\[1mm]
O (13.6 eV) & \multicolumn{2}{c}{0.65$^{+ 0.28}_{- 0.16}$} &  solar\\[1mm]
Ne  (21.6 eV) & \multicolumn{2}{c}{1.51$^{+ 0.49}_{- 0.38}$}  &solar\\[1mm] \hline\\[-3mm]
$\chi^2_{red}${\tiny(d.o.f.)} & \multicolumn{2}{c}{1.05 (432)} & \\[0.5mm]\hline\\[-3mm]
$L_{\rm X}$ {\tiny (0.2-8.0 keV)} & 1.37 (0.99)& 1.64 (1.13)&  $10^{30}$\,erg\,s$^{-1}$\\\hline\\[-2mm]
&\multicolumn{2}{c}{PN}&\\\hline\\[-3mm]
kT1 &\multicolumn{2}{c}{0.24$^{+ 0.03}_{- 0.03}$} & keV  \\[1mm]
kT2 &\multicolumn{2}{c}{0.64$^{+ 0.02}_{- 0.02}$} &  keV\\[1mm]
kT3 & \multicolumn{2}{c}{1.95$^{+ 0.11}_{- 0.10}$} &  keV \\[1mm]
EM1 & 0.5$^{+ 0.2}_{- 0.3}$& 2.0$^{+ 0.3}_{- 0.3}$&   $10^{52}$cm$^{-3}$\\[1mm]
EM2 & 3.5$^{+ 0.4}_{- 0.4}$& 5.6$^{+ 0.3}_{- 0.4}$&   $10^{52}$cm$^{-3}$\\[1mm]
EM3 & 5.9$^{+ 0.3}_{- 0.4}$  & 5.2$^{+ 0.4}_{- 0.3}$ & $10^{52}$cm$^{-3}$\\[1mm] \hline\\[-3mm]
$\chi^2_{red}${\tiny(d.o.f.)} & \multicolumn{2}{c}{1.04 (508)} & \\[0.5mm]\hline\\[-3mm]
$L_{\rm X}$ {\tiny (0.2-8.0 keV)} & 1.24 (0.90)& 1.62 (1.12)&  $10^{30}$\,erg\,s$^{-1}$\\\hline
\end{tabular}
\end{center}
\end{table}

The emission measure distributions show that intermediate ($\sim$~6\,--\,8~MK) and high ($\gtrsim$~20~MK) temperature plasma dominates the X-ray emission from DN~Tau in both observations, whereas
the cool component around 2~MK contributes only about 5\,\% (2005) and 15\,\% (2010) to the total emission measure. 
The respective contribution of the three components to the spectral model is shown by the dashed lines in Fig.\,\ref{pnspec}.
While the fitted temperatures are comparable between the 2005 and 2010 data for all plasma components, the emission measure of the individual components varies distinctly.
We find a strong EM-increase by roughly a factor three in the cool component and a moderate increase by 50\,\%
in the intermediate temperature component. In contrast, a constant EM or even moderate decrease is present in the hot component.
In relative terms the EM-increase is most pronounced in the cool component, but in absolute terms the increase in the intermediate temperature component
is at least comparable or even slightly larger.

The X-ray luminosity of $1.6 \times 10^{30}$\,erg\,s$^{-1}$ obtained for the 2010 {\it XMM-Newton} data is about 25\,\% higher than those in the {\it XMM-Newton} observation from 2005, but a factor
of three above the values obtained from {\it Einstein} data roughly 30~years ago 
and from {\it ROSAT} data in the early 1990s.
Neither the 2005 nor the 2010 exposure are dominated by strong flaring, thus significant, likely long-term, variability of DN~Tau's X-ray brightness must be present and clearly
this distinct change has to occur in the emission components associated with magnetic activity.

Our spectral modeling shows that the moderate X-ray brightening is caused by an increase in EM in the cool and intermediate plasma component;
these components contribute in 2010 much stronger to the EMD than in 2005.
Typically magnetically more active phases show harder spectra due to the stronger contribution from hotter plasma, but since cooler and
hotter coronal regions are not co-spatial and significant evolution may have occurred over the five years, a coronal origin for the EMD changes cannot be completely ruled out by the plasma temperatures alone.
Given the CTTS nature of DN~Tau and that a similar trend, albeit on timescales of hours, was observed on the prototype of an accretion dominated CTTS 
TW~Hya \citep{rob06}, another possibility would be to attribute the enhanced emission from cool plasma on DN~Tau to the presence of a stronger accretion component. 
In this scenario the coolest component would be naturally predominantly affected, since here the contribution from the accretion shock is largest.
While there is also a coronal contribution to the low temperature plasma and
clearly the 8~MK plasma does not originate directly from the accretion shocks, the EM-increase at intermediate temperatures
might be a contribution from an accretional-fed coronal component as suggested by \cite{bri10} in their study of TW~Hya.
The fact that the hot component ($\gtrsim$~20\,MK), attributed to the corona of DN~Tau, stayed approximately constant with a tendency for a mild decrease, 
does again not favor enhanced magnetic activity as origin of the increased X-ray brightness.
Further, this scenario would imply that the enhanced accretion component had at best a very moderate effect on the hot coronal structures associated with
the magnetically most active regions on the surface of DN~Tau. 

We find an overall low metallicity in the X-ray spectra of DN~Tau, however significant differences for individual elemental abundances are present.
The derived abundance pattern of DN~Tau shows in general a so-called IFIP (inverse First Ionization Potential) pattern that is commonly observed in active stars, where the low FIP elements like Fe are significantly depleted and
especially the high FIP elements like Ne are enhanced compared to solar composition.
The IFIP trend is not strictly linear in DN~Tau (see Table~\ref{specres} where the FIP of each element is given in brackets), 
but appears to be have a broad abundance minimum at low to intermediate FIP elements (Fe-Si-S) while
the very low FIP element Mg and the intermediate FIP element O have higher abundances and only Ne is enhanced compared to solar photospheric values. 
While the absolute abundances vary moderately between the applied models, the derived abundance ratios are fairly robust.
Independent of the specific model or data used, our best fits give a Ne/O ratio as well as a O/Fe ratio of roughly two for DN~Tau,
similar to values observed for BP~Tau \citep{rob06} and in many active M dwarf coronae \citep{gue01,rob05}.

\subsubsection{The spectrum beyond 6 keV}

In the inset of Fig.~\ref{pnspec} we show the X-ray emission from DN~Tau at very high energies; here
the PN spectra above 6.0~keV from the 2010 observation roughly splitted in the middle and binned to a minimum of five counts.
The comparison shows that photons at these energies were predominantly collected during the second and more active half of the observation, defined as $t > 55$~ks.
We identify probable contributions from the 6.4~keV Fe-K$\alpha$ fluorescence line, which is excited by photons with energies above 7.1~keV,
from the 6.7~keV \ion{Fe}{xxv} line complex and possibly also from the 6.97~keV \ion{Fe}{xxvi} line.
When adding a narrow Gaussian at 6.4~keV to the 2010 spectral model, where fluorescence photons were not included, we find that
the Fe-K$\alpha$ line is formally detected, but its flux is with $2.1 (0.3-3.9) \times 10^{-15}$~ergs\,cm$^{-2}$\,s$^{-1}$ poorly defined.
The additional presence of emission lines from highly ionized Fe indicates that plasma with temperatures of $\gtrsim 40$\,MK is generated in active structures on DN~Tau, 
most likely predominantly during the detected flares.
Nevertheless, while the spectra clearly suggest the presence of very hot plasma on DN~Tau, especially in the more active half of the 2010 observation,
even at this phase its contribution to the total X-ray emission is with a few percent very minor.

\subsubsection{X-ray absorption towards DN Tau}

Absorption can significantly alter the appearance of X-ray spectra and we derive from our modeling a moderate absorption column density of $N_{H}=0.8 \times 10^{21}$\,cm$^{-2}$, 
showing that no large amounts of circumstellar or disk material are in the line of sight.
The X-ray absorption is, in contrast to extinction, also sensitive to optically transparent material and thus it is a useful tool to study infalling or outflowing dust-free gas or plasma. As mentioned above, the modelled X-ray absorption is virtually unaffected by the observed changes in the EMD.
Consequently, if the cooler X-ray plasma is largely created in the vicinity of the accretion shocks and the increase in emission measure is caused by a higher mass accretion rate, 
then the plasma in the accretion columns
can have at maximum a very moderate contribution to the modeled X-ray absorption in DN~Tau. 
The X-ray absorption of DN~Tau is overall consistent with the one expected from the optical extinction $A_{\rm V} \approx 0.3 \dots 0.5$~mag, when
using the standard conversion $N_{\rm H} = 1.8 \times 10^{21}$\,cm$^{-2} \times A_{\rm V}$\,cm$^{-2}$ \citep{pred95}.
Adopting a roughly standard gas-to-dust ratio, an extinction of $A_{V} = 0.9$ as used by \cite{ing13} is not supported by the X-ray results.
Several other CTTS (e.g. BP~Tau) also show an agreement within a factor of two between X-ray and optical absorption,
in contrast the more pole-on CTTS RU~Lup \citep{rob07b} or the near edge-on system AA~Tau \citep{schmitt07} exhibit
an X-ray absorption that is up to about one magnitude above the values derived from optical measurements 
and indeed most CTTS show an excess X-ray absorption in \cite{guenther08}.
This finding indicates that mainly matter with a 'normal', i.e. roughly interstellar, gas-to-dust ratio
is responsible for the absorption towards the X-ray emitting regions on DN~Tau. Significant amounts of optically transparent material like accretion streams or hot winds are absent in the line of sight, probably favored by the fact that DN~Tau is viewed under an intermediate inclination.

\subsection{High-resolution X-ray spectroscopy}

\begin{figure}[t]
\includegraphics[width=90mm]{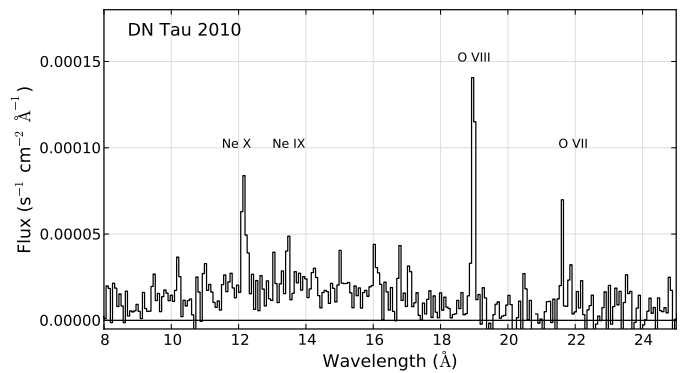}
\caption{\label{rgs} Flux converted RGS spectrum (2010 observation) of DN Tau.}
\end{figure}

The high-resolution RGS spectrum of DN~Tau obtained in 2010 is shown in Fig.\,\ref{rgs}, here flux-converted in the 8\,--\,25\,\AA\, range. 
A global modelling of these data leads to overall similar results as derived above and we concentrate in the following 
on the analysis of the brighter emission lines denoted in the plot.
These density and temperature sensitive lines are of special diagnostic interest, since they can be used to study the 
plasma contributions originating in the corona and accretion spots.

In our analysis we use the lines of the He-like triplet 
of \ion{O}{vii}, namely resonance\,(r), intercombination\,(i) and forbidden\,(f) at 21.6, 21.8, 22.1\,\AA, as well as the Ly\,$\alpha$ line of \ion{O}{viii} at 18.97\,\AA.
The absorption corrected photon fluxes, using the $N_{H}$ value from our EPIC modelling, of the relevant lines are given in Table\,\ref{lines} and
a zoom on the \ion{O}{vii} triplet for the two exposures is shown in Fig.\,\ref{o7}. We also make a comparison to the 2005 observation to check our results from the global spectroscopy, but admittedly the S/N of these data is rather poor. 
A similar diagnostic for moderately hotter plasma uses the \ion{Ne}{ix} triplet (13.45\,--\,13.7\,\AA) and the \ion{Ne}{x} line (12.1\,\AA) which are detected in the 2010 spectrum.
These lines allow also an abundance analysis of the Ne/O ratio when applying the methods described in \cite{rob08}. For DN~Tau we find Ne/O~$\approx 0.4$,
a typical value for an active star and similar to the one derived above.

\begin{table}[t]
\setlength\tabcolsep{5pt}
\caption{\label{lines}Line fluxes in $10^{-5}$~cts\,cm$^{-2}$\,s$^{-1}$, absorption corrected.}
\begin{center}{
\begin{tabular}{lrrrr}\hline\hline\\[-3mm]
Data &\multicolumn{1}{c}{Ly$\alpha$} & \multicolumn{1}{c}{r} & \multicolumn{1}{c}{i}  & \multicolumn{1}{c}{f}\\\hline\\[-3mm]
 & {OVIII} & \multicolumn{3}{c}{OVII}\\\hline\\[-3mm]
2010 & 4.0$\pm 0.5$ &  1.4$\pm$0.4 & 1.4$\pm$0.5 & 0.5$\pm$0.3 \\
2005 &2.3$\pm$1.2 &    0.7$\pm$0.5 & 1.3$\pm$0.7& 1.2$\pm$0.7 \\\hline\\[-3mm]
 & {Ne X} & \multicolumn{3}{c}{Ne IX}\\\hline\\[-3mm]
2010 & 1.6$\pm 0.2$ & 0.8$\pm$0.2 & 0.2$\pm$0.2 & 0.7$\pm$0.2 \\\hline
\end{tabular}}
\end{center}
\end{table}

\subsubsection{Oxygen lines - plasma density}
\label{o7sect}

To search for high density plasma from accretion shocks, we specifically study the
density sensitive $f/i$\,-\,ratio of the \ion{O}{vii} triplet \citep[see e.g.][]{por01}, that has a peak formation temperature of about $2$~MK. 

The plasma density is determined from the relation $ f/i =R_{0}$\,/\,$(1+\phi/\phi_{c}+n_{e}/N_{c})$ with $f$ and $i$ being the respective line intensities,
$R_{0}=3.95$ the low density limit of the line ratio,
$N_{c} =3.1 \times 10^{10}$cm$^{-3}$ the critical density and $\phi/\phi_{c}$ the radiation term.
The effect from radiation is neglected in our calculations since the UV field of DN~Tau is not sufficiently strong to influence the \ion{O}{vii} ratio.
A strong FUV flux would lower the derived plasma densities, but 
in the case of DN~Tau the FUV emission would also to be attributed to the accretion shocks, that however produce only a rather small UV-excess.
On the other hand, \ion{O}{vii} is not only produced in the accretion shocks, but also in the corona which is dominated by low density plasma
and the true accretion shock density would be underestimated.
As a consequence, changes in the measured \ion{O}{vii} density can be caused either by changes of the actual densities in the accretion components or by the relative mixture 
of low and high density plasma from the corona and the accretion shocks.

\begin{figure}[t]
\begin{center}
\includegraphics[width=85mm]{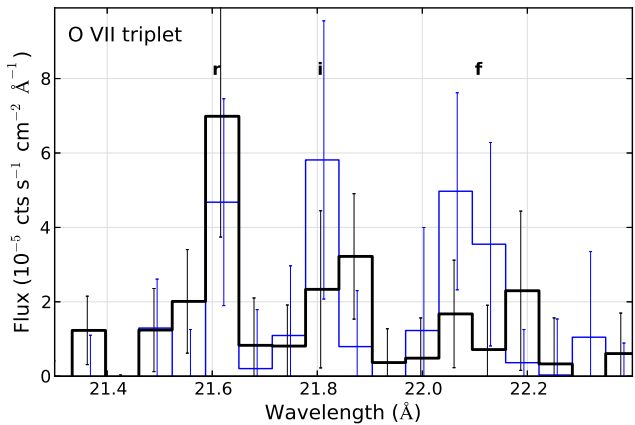}
\caption{\label{o7}Observed \ion{O}{vii} triplet in 2010 (black) and 2005 (blue).}
\end{center}
\end{figure}

As shown in Fig.\,\ref{o7}, the \ion{O}{vii} intercombination line is stronger than the forbidden line in the 2010 spectrum, while in the 2005 data
they are of comparable strength.
We find a $f/i$\,-\,ratio below one in both observations;
the derived values are $f/i = 0.36 \pm 0.26$ for the 2010 data and $f/i =0.92 \pm 0.73$ for the 2005 data from measured line fluxes.
Poissonian ranges (90\,\% conf.) derived from Monto-Carlo methods on the measured counts are 0.14\,--\,0.62 (2010) and 0.06\,--\,2.0 (2005).
Coronal sources typically exhibit a higher ratio of $f/i \gtrsim 1.5$ \citep{ness04}, 
indicating the presence of non-coronal plasma on DN~Tau.
The $f/i$\,-ratio differs by a factor of 2.5 between the observations, but large errors, especially for the 2005 exposure, are present. 
For the \ion{O}{vii} emitting plasma we find a density of $n_{e} = 3.0~(1.6-11.8) \times 10^{11}$~cm$^{-3}$ (2010) and $n_{e}=1.0~(0.4-6.1) \times 10^{11}$~cm$^{-3}$ (2005) respectively; given the coronal contribution these values are likely lower limits for the accretion shocks.
Overall the densities derived for DN~Tau are comparable, but at the lower end of values found for other CTTS.
Given a theoretical 'low density' $f/i$\,-\,ratio of around four, the coronal contribution at \ion{O}{vii} temperatures is expected to be only moderate and an
inspection of our spectral model shows that virtually all ($\sim 90\%$) of the \ion{O}{vii} emission is generated by the coolest plasma component.
An apparent higher density in 2010 would be naturally explained by a stronger contribution of accretion plasma to the X-ray emission, 
either due to a larger spatial extend or to a higher density of the shock region(s). 

Inspecting the $f/i$\,-\,ratio for high densities, i.e. $\log n_{e} \gtrsim 12/13$~cm$^{-3}$, one finds $f/i \lesssim 0.1/0.01$.
Thus the \ion{O}{vii} triplet is at the very end of its density sensitive range and
even small contributions from the omnipresent corona can strongly affect the results. 
For example, adding to a high density plasma with $\log n_{e} \sim 13$~cm$^{-3}$ a 10\,\% fraction of coronal material ($f/i \sim 3$), already
reduces the apparent density by about one order of magnitude.
In these cases the \ion{O}{vii} analysis
does not measure the true density of the accretion shock or the average density of the stellar plasma, but 
primarily traces the relative contributions from the visible portions of the accretion shocks and corona.
Assuming shocks with high density and a corona with low density, we derive an accretion to coronal EM-ratio of about 0.7:1 in 2005 that
increased to a ratio of 2:1 in the year 2010. Correspondingly one would
expect, assuming a similar corona, a comparable increase in the coolest plasma component and while there appears to be some deficit in total
\ion{O}{vii} photons, this is roughly consistent with our finding from \ion{O}{vii}($r$) and the EMD modelling. 

A density analysis of hotter plasma at $\approx$~4\,MK can be carried out with the \ion{Ne}{ix} triplet, that also has a higher critical density.
Here we find a $f/i$\,-\,ratio that is compatible with its low density limit ($n_{e} \approx 1.0\times 10^{11}$cm$^{-3}$),
while at the high end we can only put an upper limit of $n_{e} \lesssim 5 \times 10^{12}$cm$^{-3}$ on the plasma density.
In addition, the \ion{Ne}{ix} $i$ line is blended by several Fe lines with \ion{Fe}{xix} being the strongest one;
this is not taken into account in above calculation since Fe is heavily depleted in DN~Tau and statistical errors on the only weakly detected $i$-line dominate. 
Moderately cooler plasma at $\sim$~1.5\,MK could be studied with the \ion{N}{vi} triplet, but the data quality is insufficient to provide any further constraints.

\subsubsection{Oxygen lines - plasma temperatures}

Accretion processes that contribute to the X-ray emission can also be studied with temperature diagnostics and
here we search for an excess of cool plasma via the \ion{O}{viii}/\ion{O}{vii} line ratio.
The material that is accreted by CTTS has infall velocities of a few hundred km\,s$^{-1}$ and thus the post-shock plasma reaches at maximum temperatures up to a few MK.
Such plasma is still relatively cool with respect to the coronal temperatures in active stars or CTTS and
should therefore be detectable as 'soft excess' X-ray emission.

As temperature diagnostic we use strong oxygen emission lines as measured in 2010, here the
\ion{O}{viii} Ly$_{\alpha}$ line (18.97~\AA) and the \ion{O}{vii} He-like triplet lines with peak formation temperatures of $\sim$~3\,MK and $\sim$~2\,MK respectively. 
An abundance independent method is obtained by using the \ion{O}{viii}(Ly$_{\alpha}$)/\ion{O}{vii}(r) energy flux ratio in comparison to the summed luminosity of both lines. 

\begin{figure}[t]
\includegraphics[width=99mm]{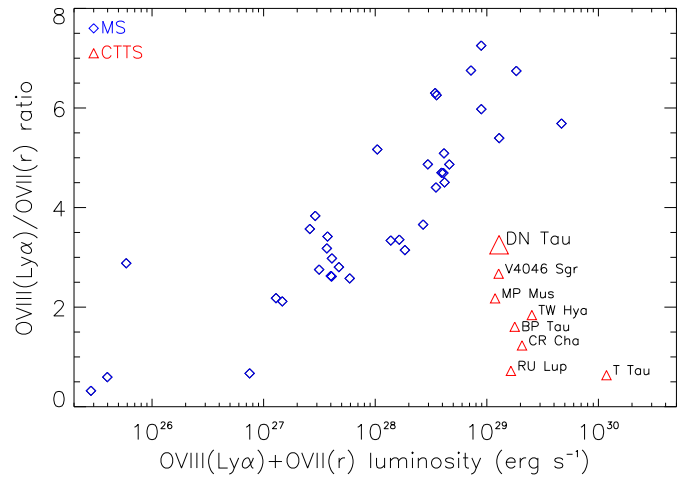}
\caption{\label{coolex} The soft excess of DN~Tau; \ion{O}{viii}(Ly$_{\alpha}$)/\ion{O}{vii}(r)-ratio vs. summed luminosity
for main-sequence stars (diamonds) and CTTS (triangles).}
\end{figure}

In Fig.\,\ref{coolex} we compare the  \ion{O}{viii}/\ion{O}{vii}-ratio of DN~Tau with those of other CTTS collected in the literature and with a large sample of main-sequence stars at various activity levels taken from \cite{ness04}.
The correlation between the \ion{O}{viii}/\ion{O}{vii} line ratio and $L_{\rm X}$ for main-sequence stars is well known and 
caused by the higher coronal temperatures in more active and X-ray brighter stars.
As shown in the plot, DN~Tau exhibits a soft excess compared to active coronal sources with similar X-ray luminosity, but it is quite weak when compared to other CTTS.
Actually the soft excess of DN~Tau is the weakest in the sample of X-ray studied CTTS.
Calculating the \ion{O}{viii}/\ion{O}{vii} energy flux ratio of DN~Tau we find a value of around three; inspecting theoretical ratios as calculated with e.g. the {\it Chianti} code this corresponds to an average plasma temperature of 3.0\,--\,3.5~MK.
This temperature is rather high given the expected shock temperatures and supports the idea that most of the oxygen emission in DN~Tau 
is produced by magnetic activity or consists of mixed accreted and coronal plasma.
Another method based on \ion{O}{vii} alone and thus more suited for very cool temperatures
uses the temperature sensitive g-ratio, $g= (f+i) / r$. Our value of $g = 1.36 \pm 0.57$ favors indeed
low temperatures of $\lesssim 1$~MK for the \ion{O}{vii} plasma, but due to its weaker temperature-dependence also twice as high temperatures are consistent with
the data.

\section{Discussion}
\label{comp}

\subsection{The accretion shocks on DN Tau}

At first look the soft excess of DN~Tau is surprisingly weak for a young CTTS that is accreting matter from its disk and that exhibits a high plasma density in \ion{O}{vii}.
Two main factors might be responsible for this effect; either the relative accretion luminosity is very low or the accreted plasma 
is not heated sufficiently to produce strong \ion{O}{vii} emission.

The estimated mass accretion rates from optical/UV observations of DN~Tau are intermediate for CTTS,
therefore they alone cannot explain 
the weakness of its soft excess. Here the ratio of coronal to accretion luminosity is another important measure.
While the density analysis suggests a significant contribution of the accretion shock to the \ion{O}{vii} emission, the relative contribution from very cool plasma to the overall X-ray emission is quite weak
and a strong \ion{O}{viii} line from the corona reduces the soft excess. 
Also the evolutionary phase plays an important role and DN~Tau is a low-mass CTTS that is relatively young and thus still quite enlarged. 
As a consequence, shock speeds and temperatures do not reach values found for more massive or older, more compact stars.
Basically, $V_{sh} \propto \sqrt{2 G M_{*}/R_{*} \times (1-R_{*}/R_{t}) }$ and $T_{sh} \propto  V_{sh}^{2}$ \citep[e.g.][]{lam98,cal98}, 
where $M_{*}$ and $R_{*}$ are stellar mass and radius and $R_{t}$ is the disk truncation radius, i.e. from where matter is falling onto the star.
The calculated shock velocity depends slightly on the adopted stellar model, using the \cite{donati13} values of $M_{*}= 0.65~M_{\odot}$ and $R_{*}=1.9~R_{\odot}$ 
and their $R_{mag} = 5.9~R_{*}$ (2010) as truncation radius,
results in shock speeds of about $V_{sh} = 330 \pm 30$~km\,s$^{-1}$.
Adopting the slightly larger or less massive stellar models with e.g. $M_{*} \approx 0.5~M_{\odot}$ and $R_{*} \approx 2.1~R_{\odot}$, 
gives $V_{sh}=$\,260\,--\,300~km\,s$^{-1}$; here we assumed accretion from the respective co-rotating radius.
The corresponding strong-shock temperatures are in the range of $0.9 - 1.5$~MK and similar to the \ion{O}{vii}-temperature derived above.
While these temperatures are sufficient to produce \ion{O}{vii} emission, they are below the peak emissivity temperature of about 2~MK, 
reducing the contribution from the accretion shocks to these lines and consequently the strength of the soft excess.
Since the hot and active corona emits predominantly at higher temperatures, it contributes even weaker to \ion{O}{vii} and
thus preserves the accretion shock signatures in the applied emission line diagnostic, most prominently seen in the very cool plasma during DN~Tau's 2010 soft state. 
However, compared to other CTTS, the 'pure' accretion shock plasma is on DN~Tau a weak contributor to the total X-ray emission.

The X-ray mass accretion rate for DN~Tau can be calculated from mass conversation under the assumption of a strong shock by using
$\dot{M}_{acc} = 4 \pi f R^{2}_{*} \rho_{pre} V_{sh}$. 
Using our best-fit modeling results on the plasma density, adopting a mean molecular weight of $\mu_{e} = 1.2$, a filling factor of $f = 0.01$  and stellar models as above,
we derive an X-ray mass accretion rate of $\log \dot{M} \approx -9.5~M_{\odot}$\,yr$^{-1}$. As a caveat and
recalling the discussion above, due to the coronal blend the \ion{O}{vii} density is likely a lower limit for the accretion shock density and
further the filling factor is adopted from other analysis.
An independent estimate based on X-ray data can be derived when fitting our X-ray spectra with the model from \cite{guenther07}, where
we obtain a mass accretion rates around $\log \dot{M} =-9.2~M_{\odot}$\,yr$^{-1}$.
With these models we find filling factors of $f \lesssim 0.01$ and post-shock densities of
$n_{e} \gtrsim 3 \times 10^{11}$~cm$^{-3}$, but their interdependency does not allow to further constrain the accretion shock properties.
The X-ray accretion rate is about one order of magnitude below the values mostly found from optical/UV measurements, similar to
result obtained for other CTTS \citep[e.g. BP~Tau,][]{schmitt05}.
While intrinsic variability likely also plays a role, this finding might indicate that the accreted material contributes only fractionally to the observed X-ray emission.

In these scenarios, either not all accreted material produces X-rays or the X-rays are produced but partially absorbed or both.
In some accretion regions the shock temperatures might be too low to generate X-ray emission and thus X-rays would trace only the fastest fraction of the accretion stream;
although it remains unclear why infalling material should not impact the stellar surface with similar velocities when accreted from similar distances, i.e. around the disk truncation radius.
Similarly, accretion streams of low density that produce no strong X-ray emission may remain mostly undetected and lead to missing material, 
but if they exist they are expected to carry only a small fraction of the total mass flux. 
For example, adding low flux columns ($F \propto \rho V^{3}$) to the model of DN~Tau, 
increases the maximum spot size ($f$) by a factor of 30, but the mass accretion rate ($\dot{M}$) by less than a factor of two \citep{ing13}.
Alternatively, virtually all infalling material might produce X-ray emission in accretion shocks, 
but these X-rays are partly absorbed locally, e.g. by the accretion column and thus missing in the observed X-ray spectra as suggested by \cite{sac10}.

Recently, \cite{dod13} performed a non-LTE modeling of emission components of optical He and Ca lines by adding an accretion hot spot and a photosphere.
Significant variability is present in the derived
accretion parameters for two DN~Tau observations performed in autumn 2009 and spring 2010.
They find pre-shock (infall) densities of $\log n_{e} =12.2/13$~cm$^{-3}$, velocities of $V_{0}= 230/280$~km\,s$^{-1}$ and filling factors of $f = 3/1.2$\,\%, leading to
accretion luminosities of 3\,/\,16\,\%~$L_{*}$ and mass accretion rates of $\log \dot{M} = -8.1 /-7.6~M_{\odot}$\,yr$^{-1}$.
These results again indicate a very high infall density, large mass accretion rates and large filling factors for DN~Tau. Beside, they support
an active accretion period in 2010, at least a few month before the new X-ray data was taken. where a strong accretion stream impacts high latitude regions. 
The 2010 data presented in \cite{donati13} was obtained a few month after 
our X-ray observation and albeit they obtain a lower mass accretion rate of $\log \dot{M} = -9.1$, their surface maps
show a large monolithic dark spot and an embedded accretion region at similar high latitudes. 
Their dominant spot is located in 2010 at about phase 0.55 and our X-ray data was taken at phase 0.1\,--\,0.3.
Thus if this configurations is applicable, it implicates a relatively unspoiled view on the accretion spot region during the {\it XMM-Newton} observation.

In summary, the low infall velocities and the non-negligible coronal contribution likely make 
X-ray diagnostics less favorable for a quantitative analysis, but they are still applicable to detect the presence of X-rays from accretion shocks in CTTS like DN Tau.

\subsection{X-ray variability}

The observed X-ray variability can be caused by intrinsic changes in accretion rate or magnetic activity as well as by varying viewing geometry, absorption and rotation.
While for the omnipresent short- and mid-term variability (seconds to months) all these factors contribute, the situation is less clear for the major cause of possible long-term trends on timescales
of years to decades.
The changes in X-ray brightness between 2005 and 2010 can likely be attributed to different accretion states, unless a large fraction of the accretion spots is 'hidden' in the 2005 exposure.
Here, variable mass accretion rates and changing magnetic topology, which influences the disk truncation radius and viewing geometry, likely play the major role.
Since the X-rays from the accretion shocks experience absorption by the above accretion streams, some time dependent viewing geometry effects may be present 
in the detected X-ray emission, as suggested in \cite{arg11} for their V2129~Oph data.
Our 2010 observation covers 0.22 in rotational phase and the line of sight is likely not aligned with the accretion stream, but the 2005 data (0.06 phase coverage) could be more severely affected.
While the global $N_{H}$ is identical for both observations and the apparent presence of high density plasma in the 2005 spectrum as well as 
the large EM changes over a broad temperature range do not strongly support this explanation for the case of DN~Tau, 
the presence of a variable contribution from locally produced and re-absorbed accretion components is not ruled out by the data.

In contrast, the increase in X-ray brightness by a factor of three compared to the measurements in the 1980s and 1990s seems to favor a change in the magnetic activity state of DN~Tau,
already because its X-ray emission detected by {\it XMM-Newton} is predominantly of coronal origin.
On the other hand, DN~Tau showed on average a brightening over the last decades in the U band, again by a factor of three, likely attributable
predominantly to the accretion shocks. These apparently different scenarios might at least partially be explained by an accretion fed corona that
would lead to the observed trends, but unfortunately most of these observations are not simultaneous and especially X-ray data is sparse.
Thus there might be no overall long-term trend at all, but instead UV bright and X-ray bright phases that are not necessarily related to each other 
and dominated by the various kinds of short-term variability.

\subsection{DN Tau in the CTTS context}

The young low-mass CTTS DN~Tau shows a soft excess and a high density in its cool ($\sim$~2\,MK) plasma component.
Both findings indicate, that plasma originating in well funnelled accretion streams impacting the stellar surface contributes
to the observed X-ray emission. The plasma densities of DN~Tau derived from \ion{O}{vii} diagnostics are similar to values of other CTTS like BP~Tau or RU~Lup \citep{schmitt05,rob07b}, but
its soft X-ray excess is quite weak for a young CTTS. The higher mass accretion rates of BP~Tau or RU~Lup naturally lead to a more pronounced soft excess, but this effect cannot
account for all CTTS. Additionally, as a consequence of the low mass and large radius of DN~Tau,
the impact velocity is among the lowest of all studied CTTS and expected accretion shock temperatures are well below the peak formation temperature
of the studied X-ray lines.
Therefore only the hottest part of the accreted and shocked material will reach X-ray temperatures, however this is still sufficient to produce detectable signatures in the X-ray data.
Combining the only moderate mass accretion rate and the low impact velocity
would explain, why the soft excess of DN~Tau is even smaller than those of old CTTS like V4046~Sgr or MP~Mus.
While their accretion rates are even lower, these objects are more massive and especially more compact ($M_{*}/R_{*}$) and the accretion shocks have higher temperatures
and produce X-ray emission in the \ion{O}{vii}-lines more efficiently.

Comparing DN~Tau to other young stars in the lower mass regime, we find that a moderate soft excess is also present in the TWA member Hen~3-600 \citep{huene07}, a multiple system with an M3/M3.5 binary as principle components.
The value of its \ion{O}{viii}/\ion{O}{vii} ratio is similar to the one of DN~Tau, albeit Hen~3-600 is about a factor five fainter in X-rays and its soft excess 
is even less pronounced.
In contrast to DN~Tau, Hen~3-600 is old ($\sim 10$~Myr) and likely already in the CTTS/WTTS transitional phase.
While expected to be compact, here the evolved state and corresponding very low accretion rate reduces accretion shock signatures in its X-ray spectrum. 
Correspondingly, its oxygen $f/i$\,-ratio of about 1.1, yet with significant error, is at the uppermost end of the values observed for accretional sources.

Notably, coronal properties as derived from the modelling of global X-ray spectra are very similar for DN~Tau and BP~Tau \citep{rob06}. We find similar X-ray luminosities and plasma temperatures
as well as nearly identical abundance patterns. The largest differences are found in the relative strength of the cool component of their EMDs, but as outlined above 
this is where the accretion shocks have their largest impact in the X-ray spectra. Overall, at least when considering stars with comparable X-ray activity,
the coronal properties seem to vary at best moderately in the regime of young low-mass CTTS when going to less massive stars.

With DN~Tau we extend the X-ray studied sample of young accreting stars to lower masses and its X-ray properties clearly link it to more massive or more evolved CTTS.
The combination of a very cool accretion component with a strong hot corona makes DN~Tau one of the X-ray brightest CTTS in its mass range, 
but reduces the influence of the accretion shocks in its X-ray spectrum and emission line diagnostics.

\section{Summary \& conclusions}
\label{sum}
From our study of the X-ray emission of DN~Tau we obtain the following main results and draw the subsequent conclusions:

\begin{enumerate}
\item DN~Tau is among the least massive CTTS where cool MK-temperature plasma at high density from accretion shocks is clearly present and
it is the youngest star in the regime of M-type stars studied in X-rays in greater detail.
DN~Tau shares general properties with other low-mass CTTS, but differences arise in detail that are mainly related to its youth and low mass.

\item The \ion{O}{vii} triplet shows an $f/i$\,-\,ratio of about 0.4, attributed to accretion shocks that significantly
contribute to the soft X-ray emission.
The corresponding plasma density is $n_{e} = 3 - 4 \times 10^{11}$~cm$^{-3}$, due to the coronal contribution this is likely a lower limit
for the accretion shocks. DN~Tau shows a soft excess as measured in \ion{O}{viii}/\ion{O}{vii}-ratios, 
confirming the presence of accretion shock plasma.
While the plasma density is quite typical when compared to other CTTS, the soft X-ray excess is rather weak. Here the low impact velocity
of the accreted material, a consequence of the low mass
and large radius of DN~Tau, results in shock temperatures of about $1.0 - 1.5$~MK, well below peak formation of \ion{O}{vii}. Overall the cool plasma component around 2~MK contributes only moderately to the X-ray emission.

\item A strong coronal component with hot ($\gtrsim$ 10 MK) plasma is present and at higher energies the spectrum of DN Tau is dominated by magnetic activity. 
Intermediate temperature plasma clearly originates from coronal structures, but may contain accretion-fed material.
The corona reaches temperatures of $\gtrsim 30$~MK and its abundance pattern shows an IFIP effect that is reminiscent of those of active stars.
DN~Tau is with $\log L_{\rm X} = 30.2$\,erg\,s$^{-1}$ among the X-ray brighter CTTS in its mass- or $T_{\rm eff}$\,-\,range.

\item We find significant changes of DN~Tau's X-ray properties; in 2010 it was in an X-ray brighter, but overall softer spectral state compared to 2005.
The emission measure of the cool plasma changed by a factor of a few, indicating accretion related variability.
Similar, but less pronounced changes are observed at intermediate temperatures; in contrast
the hot component stayed virtually constant. No changes in absorption column or elemental abundances were found; further the X-ray absorption is consistent with optical values.

\item Several X-ray flares with durations of $\lesssim 1$~h are detected in the 2010 exposure, which are accompanied by clear UV counterparts. 
The UV emission precedes the X-rays as expected in the chromospheric evaporation scenario and the flares are similar to the ones seen on active young M dwarfs.
Outside the observed flares an obvious correlation between X-ray and UV brightness is not observed, indicating 
largely independent emission regions. Brightness differences by a factor of three are present in X-ray and U band data on timescales of years to decades. 

\end{enumerate}

\begin{acknowledgements}
This work is based on observations obtained with {\it XMM-Newton},
an ESA science mission with instruments and contributions directly
funded by ESA Member States and the USA (NASA).
J.R. acknowledges support from the DLR under grant 50QR0803. 
HMG was supported by the National Aeronautics and Space Administration
under Grant No. NNX11AD12G issued through the Astrophysics Data Analysis Program.
The publication is supported by the Austrian Science Fund (FWF).
\end{acknowledgements}

\bibliographystyle{aa}
\bibliography{dntau.bbl}

\begin{thebibliography}{59}
\expandafter\ifx\csname natexlab\endcsname\relax\def\natexlab#1{#1}\fi

\bibitem[{{Andrews} \& {Williams}(2005)}]{and05}
{Andrews}, S.~M. \& {Williams}, J.~P. 2005, \apj, 631, 1134

\bibitem[{{Argiroffi} {et~al.}(2011){Argiroffi}, {Flaccomio}, {Bouvier},
  {Donati}, {Getman}, {Gregory}, {Hussain}, {Jardine}, {Skelly}, \&
  {Walter}}]{arg11}
{Argiroffi}, C., {Flaccomio}, E., {Bouvier}, J., {et~al.} 2011, \aap, 530, A1

\bibitem[{{Argiroffi} {et~al.}(2012){Argiroffi}, {Maggio}, {Montmerle},
  {Huenemoerder}, {Alecian}, {Audard}, {Bouvier}, {Damiani}, {Donati},
  {Gregory}, {G{\"u}del}, {Hussain}, {Kastner}, \& {Sacco}}]{arg12}
{Argiroffi}, C., {Maggio}, A., {Montmerle}, T., {et~al.} 2012, \apj, 752, 100

\bibitem[{{Argiroffi} {et~al.}(2007){Argiroffi}, {Maggio}, \& {Peres}}]{arg07}
{Argiroffi}, C., {Maggio}, A., \& {Peres}, G. 2007, \aap, 465, L5

\bibitem[{{Arnaud}(1996)}]{xspec}
{Arnaud}, K.~A. 1996, in ASP Conf.\ Ser, Vol. 101, Astronomical Data Analysis
  Software and Systems V, ed. G.~H. {Jacoby} \& J.~{Barnes} (San Francisco:
  ASP), 17

\bibitem[{{Bouvier} {et~al.}(1986){Bouvier}, {Bertout}, \& {Bouchet}}]{bou86}
{Bouvier}, J., {Bertout}, C., \& {Bouchet}, P. 1986, \aap, 158, 149

\bibitem[{{Brickhouse} {et~al.}(2010){Brickhouse}, {Cranmer}, {Dupree}, {Luna},
  \& {Wolk}}]{bri10}
{Brickhouse}, N.~S., {Cranmer}, S.~R., {Dupree}, A.~K., {Luna}, G.~J.~M., \&
  {Wolk}, S. 2010, \apj, 710, 1835

\bibitem[{{Calvet} \& {Gullbring}(1998)}]{cal98}
{Calvet}, N. \& {Gullbring}, E. 1998, \apj, 509, 802

\bibitem[{{Cohen} \& {Kuhi}(1979)}]{coh79}
{Cohen}, M. \& {Kuhi}, L.~V. 1979, \apjs, 41, 743

\bibitem[{{de la Calle}(2012)}]{sas}
{de la Calle}, I. 2012, http://xmm.esac.esa.int

\bibitem[{{Dodin} {et~al.}(2013){Dodin}, {Lamzin}, \& {Sitnova}}]{dod13}
{Dodin}, A.~V., {Lamzin}, S.~A., \& {Sitnova}, T.~M. 2013, Astronomy Letters,
  39, 315

\bibitem[{{Donati} {et~al.}(2013){Donati}, {Gregory}, {Alencar}, {Hussain},
  {Bouvier}, {Jardine}, {M{\'e}nard}, {Dougados}, {Romanova}, \& {MaPP
  collaboration}}]{donati13}
{Donati}, J.-F., {Gregory}, S.~G., {Alencar}, S.~H.~P., {et~al.} 2013, \mnras,
  436, 881

\bibitem[{{Fernandez} {et~al.}(1995){Fernandez}, {Ortiz}, {Eiroa}, \&
  {Miranda}}]{fern95}
{Fernandez}, M., {Ortiz}, E., {Eiroa}, C., \& {Miranda}, L.~F. 1995, \aaps,
  114, 439

\bibitem[{{Furlan} {et~al.}(2011){Furlan}, {Luhman}, {Espaillat}, {D'Alessio},
  {Adame}, {Manoj}, {Kim}, {Watson}, {Forrest}, {McClure}, {Calvet}, {Sargent},
  {Green}, \& {Fischer}}]{fur11}
{Furlan}, E., {Luhman}, K.~L., {Espaillat}, C., {et~al.} 2011, \apjs, 195, 3

\bibitem[{{Getman} {et~al.}(2005){Getman}, {Flaccomio}, {Broos}, {Grosso},
  {Tsujimoto}, {Townsley}, {Garmire}, {Kastner}, {Li}, {Harnden}, {Wolk},
  {Murray}, {Lada}, {Muench}, {McCaughrean}, {Meeus}, {Damiani}, {Micela},
  {Sciortino}, {Bally}, {Hillenbrand}, {Herbst}, {Preibisch}, \&
  {Feigelson}}]{get05}
{Getman}, K.~V., {Flaccomio}, E., {Broos}, P.~S., {et~al.} 2005, \apjs, 160,
  319

\bibitem[{{Grankin} {et~al.}(2007){Grankin}, {Melnikov}, {Bouvier}, {Herbst},
  \& {Shevchenko}}]{gran07}
{Grankin}, K.~N., {Melnikov}, S.~Y., {Bouvier}, J., {Herbst}, W., \&
  {Shevchenko}, V.~S. 2007, \aap, 461, 183

\bibitem[{{Grevesse} \& {Sauval}(1998)}]{grsa}
{Grevesse}, N. \& {Sauval}, A.~J. 1998, Space Science Reviews, 85, 161

\bibitem[{{G{\"u}del} {et~al.}(2001){G{\"u}del}, {Audard}, {Magee},
  {Franciosini}, {Grosso}, {Cordova}, {Pallavicini}, \& {Mewe}}]{gue01}
{G{\"u}del}, M., {Audard}, M., {Magee}, H., {et~al.} 2001, \aap, 365, L344

\bibitem[{{G{\"u}del} {et~al.}(2007{\natexlab{a}}){G{\"u}del}, {Briggs},
  {Arzner}, {Audard}, {Bouvier}, {Feigelson}, {Franciosini}, {Glauser},
  {Grosso}, {Micela}, {Monin}, {Montmerle}, {Padgett}, {Palla}, {Pillitteri},
  {Rebull}, {Scelsi}, {Silva}, {Skinner}, {Stelzer}, \& {Telleschi}}]{gue07}
{G{\"u}del}, M., {Briggs}, K.~R., {Arzner}, K., {et~al.} 2007{\natexlab{a}},
  \aap, 468, 353

\bibitem[{{G{\"u}del} {et~al.}(2005){G{\"u}del}, {Skinner}, {Briggs}, {Audard},
  {Arzner}, \& {Telleschi}}]{gue05}
{G{\"u}del}, M., {Skinner}, S.~L., {Briggs}, K.~R., {et~al.} 2005, \apjl, 626,
  L53

\bibitem[{{G{\"u}del} {et~al.}(2007{\natexlab{b}}){G{\"u}del}, {Skinner},
  {Mel'Nikov}, {Audard}, {Telleschi}, \& {Briggs}}]{gue07a}
{G{\"u}del}, M., {Skinner}, S.~L., {Mel'Nikov}, S.~Y., {et~al.}
  2007{\natexlab{b}}, \aap, 468, 529

\bibitem[{{G{\"u}del} \& {Telleschi}(2007)}]{gue07b}
{G{\"u}del}, M. \& {Telleschi}, A. 2007, \aap, 474, L25

\bibitem[{{Gullbring} {et~al.}(1998){Gullbring}, {Hartmann}, {Briceno}, \&
  {Calvet}}]{gull98}
{Gullbring}, E., {Hartmann}, L., {Briceno}, C., \& {Calvet}, N. 1998, \apj,
  492, 323

\bibitem[{{G{\"u}nther} {et~al.}(2006){G{\"u}nther}, {Liefke}, {Schmitt},
  {Robrade}, \& {Ness}}]{guenther06}
{G{\"u}nther}, H.~M., {Liefke}, C., {Schmitt}, J.~H.~M.~M., {Robrade}, J., \&
  {Ness}, J.-U. 2006, \aap, 459, L29

\bibitem[{{G{\"u}nther} \& {Schmitt}(2008)}]{guenther08}
{G{\"u}nther}, H.~M. \& {Schmitt}, J.~H.~M.~M. 2008, \aap, 481, 735

\bibitem[{{G{\"u}nther} {et~al.}(2007){G{\"u}nther}, {Schmitt}, {Robrade}, \&
  {Liefke}}]{guenther07}
{G{\"u}nther}, H.~M., {Schmitt}, J.~H.~M.~M., {Robrade}, J., \& {Liefke}, C.
  2007, \aap, 466, 1111

\bibitem[{{Hartmann} \& {Stauffer}(1989)}]{hart89}
{Hartmann}, L. \& {Stauffer}, J.~R. 1989, \aj, 97, 873

\bibitem[{{Herbig} \& {Bell}(1988)}]{herb88}
{Herbig}, G.~H. \& {Bell}, K.~R. 1988, {Third Catalog of Emission-Line Stars of
  the Orion Population : 3 : 1988}

\bibitem[{{Huenemoerder} {et~al.}(2007){Huenemoerder}, {Kastner}, {Testa},
  {Schulz}, \& {Weintraub}}]{huene07}
{Huenemoerder}, D.~P., {Kastner}, J.~H., {Testa}, P., {Schulz}, N.~S., \&
  {Weintraub}, D.~A. 2007, \apj, 671, 592

\bibitem[{{Ingleby} {et~al.}(2013){Ingleby}, {Calvet}, {Herczeg}, {Blaty},
  {Walter}, {Ardila}, {Alexander}, {Edwards}, {Espaillat}, {Gregory},
  {Hillenbrand}, \& {Brown}}]{ing13}
{Ingleby}, L., {Calvet}, N., {Herczeg}, G., {et~al.} 2013, \apj, 767, 112

\bibitem[{{Johns-Krull}(2007)}]{joh07}
{Johns-Krull}, C.~M. 2007, \apj, 664, 975

\bibitem[{{Kashyap} \& {Drake}(2000)}]{poa}
{Kashyap}, V. \& {Drake}, J.~J. 2000, Bulletin of the Astronomical Society of
  India, 28, 475

\bibitem[{{Kastner} {et~al.}(2002){Kastner}, {Huenemoerder}, {Schulz},
  {Canizares}, \& {Weintraub}}]{kas02}
{Kastner}, J.~H., {Huenemoerder}, D.~P., {Schulz}, N.~S., {Canizares}, C.~R.,
  \& {Weintraub}, D.~A. 2002, \apj, 567, 434

\bibitem[{{Kenyon} \& {Hartmann}(1995)}]{ken95}
{Kenyon}, S.~J. \& {Hartmann}, L. 1995, \apjs, 101, 117

\bibitem[{{Koenigl}(1991)}]{koe91}
{Koenigl}, A. 1991, \apjl, 370, L39

\bibitem[{{Lamzin}(1998)}]{lam98}
{Lamzin}, S.~A. 1998, Astronomy Reports, 42, 322

\bibitem[{{Mitra-Kraev} {et~al.}(2005){Mitra-Kraev}, {Harra}, {G{\"u}del},
  {Audard}, {Branduardi-Raymont}, {Kay}, {Mewe}, {Raassen}, \& {van
  Driel-Gesztelyi}}]{mit05}
{Mitra-Kraev}, U., {Harra}, L.~K., {G{\"u}del}, M., {et~al.} 2005, \aap, 431,
  679

\bibitem[{{Muzerolle} {et~al.}(2003){Muzerolle}, {Calvet}, {Hartmann}, \&
  {D'Alessio}}]{muz03}
{Muzerolle}, J., {Calvet}, N., {Hartmann}, L., \& {D'Alessio}, P. 2003, \apjl,
  597, L149

\bibitem[{{Ness} {et~al.}(2004){Ness}, {G{\"u}del}, {Schmitt}, {Audard}, \&
  {Telleschi}}]{ness04}
{Ness}, J.-U., {G{\"u}del}, M., {Schmitt}, J.~H.~M.~M., {Audard}, M., \&
  {Telleschi}, A. 2004, \aap, 427, 667

\bibitem[{{Ness} \& {Wichmann}(2002)}]{cora}
{Ness}, J.-U. \& {Wichmann}, R. 2002, Astron. Nachr., 323, 129

\bibitem[{{Neuh{\"a}user} {et~al.}(1995){Neuh{\"a}user}, {Sterzik}, {Schmitt},
  {Wichmann}, \& {Krautter}}]{neu95}
{Neuh{\"a}user}, R., {Sterzik}, M.~F., {Schmitt}, J.~H.~M.~M., {Wichmann}, R.,
  \& {Krautter}, J. 1995, \aap, 297, 391

\bibitem[{{Nguyen} {et~al.}(2009){Nguyen}, {Jayawardhana}, {van Kerkwijk},
  {Brandeker}, {Scholz}, \& {Damjanov}}]{ngu09}
{Nguyen}, D.~C., {Jayawardhana}, R., {van Kerkwijk}, M.~H., {et~al.} 2009,
  \apj, 695, 1648

\bibitem[{{Porquet} {et~al.}(2001){Porquet}, {Mewe}, {Dubau}, {Raassen}, \&
  {Kaastra}}]{por01}
{Porquet}, D., {Mewe}, R., {Dubau}, J., {Raassen}, A.~J.~J., \& {Kaastra},
  J.~S. 2001, \aap, 376, 1113

\bibitem[{{Predehl} \& {Schmitt}(1995)}]{pred95}
{Predehl}, P. \& {Schmitt}, J.~H.~M.~M. 1995, \aap, 293, 889

\bibitem[{{Robrade} \& {Schmitt}(2005)}]{rob05}
{Robrade}, J. \& {Schmitt}, J.~H.~M.~M. 2005, \aap, 435, 1073

\bibitem[{{Robrade} \& {Schmitt}(2006)}]{rob06}
{Robrade}, J. \& {Schmitt}, J.~H.~M.~M. 2006, \aap, 449, 737

\bibitem[{{Robrade} \& {Schmitt}(2007)}]{rob07b}
{Robrade}, J. \& {Schmitt}, J.~H.~M.~M. 2007, \aap, 473, 229

\bibitem[{{Robrade} {et~al.}(2008){Robrade}, {Schmitt}, \& {Favata}}]{rob08}
{Robrade}, J., {Schmitt}, J.~H.~M.~M., \& {Favata}, F. 2008, \aap, 486, 995

\bibitem[{{Sacco} {et~al.}(2010){Sacco}, {Orlando}, {Argiroffi}, {Maggio},
  {Peres}, {Reale}, \& {Curran}}]{sac10}
{Sacco}, G.~G., {Orlando}, S., {Argiroffi}, C., {et~al.} 2010, \aap, 522, A55

\bibitem[{{Schmitt}(1997)}]{schmitt97}
{Schmitt}, J.~H.~M.~M. 1997, \aap, 318, 215

\bibitem[{{Schmitt} \& {Robrade}(2007)}]{schmitt07}
{Schmitt}, J.~H.~M.~M. \& {Robrade}, J. 2007, \aap, 462, L41

\bibitem[{{Schmitt} {et~al.}(2005){Schmitt}, {Robrade}, {Ness}, {Favata}, \&
  {Stelzer}}]{schmitt05}
{Schmitt}, J.~H.~M.~M., {Robrade}, J., {Ness}, J.-U., {Favata}, F., \&
  {Stelzer}, B. 2005, \aap, 432, L35

\bibitem[{{Smith} {et~al.}(2001){Smith}, {Brickhouse}, {Liedahl}, \&
  {Raymond}}]{apec}
{Smith}, R.~K., {Brickhouse}, N.~S., {Liedahl}, D.~A., \& {Raymond}, J.~C.
  2001, \apjl, 556, L91

\bibitem[{{Stelzer} \& {Schmitt}(2004)}]{ste04}
{Stelzer}, B. \& {Schmitt}, J.~H.~M.~M. 2004, \aap, 418, 687

\bibitem[{{Telleschi} {et~al.}(2007){Telleschi}, {G{\"u}del}, {Briggs},
  {Audard}, \& {Scelsi}}]{tel07}
{Telleschi}, A., {G{\"u}del}, M., {Briggs}, K.~R., {Audard}, M., \& {Scelsi},
  L. 2007, \aap, 468, 443

\bibitem[{{Vrba} {et~al.}(1993){Vrba}, {Chugainov}, {Weaver}, \&
  {Stauffer}}]{vrba93}
{Vrba}, F.~J., {Chugainov}, P.~F., {Weaver}, W.~B., \& {Stauffer}, J.~S. 1993,
  \aj, 106, 1608

\bibitem[{{Walter} \& {Kuhi}(1981)}]{wal81}
{Walter}, F.~M. \& {Kuhi}, L.~V. 1981, \apj, 250, 254

\bibitem[{{White} \& {Basri}(2003)}]{whi03}
{White}, R.~J. \& {Basri}, G. 2003, \apj, 582, 1109

\bibitem[{{White} \& {Hillenbrand}(2004)}]{whi04}
{White}, R.~J. \& {Hillenbrand}, L.~A. 2004, \apj, 616, 998

\end{thebibliography}

\end{document}